\documentclass[aps,pre,showpacs,twocolumn]{revtex4}
\usepackage{bm}
\usepackage{graphicx}
\bibstyle{approve.bib}
\usepackage{amssymb}
\usepackage{amsmath}
\usepackage{esint}
\usepackage{epstopdf}
\usepackage{color}
\usepackage{gensymb}
\usepackage{mathtools}
\usepackage{suffix}
\usepackage{fancyhdr}

\newcommand{\be}{\begin{equation}}
\newcommand{\ee}{\end{equation}}
\newcommand{\bea}{\begin{eqnarray}}
\newcommand{\eea}{\end{eqnarray}}

\newcommand{\lan}{\left\langle}
\newcommand{\ran}{\right\rangle}
\newcommand{\br}{\mathbf{r}}

\newcommand{\hk}{\hat{k}}

\newcommand{\bk}{\mathbf{k}}

\newcommand{\e}{\varepsilon}

\newcommand{\tv}{\tilde{v}}

\newcommand{\tP}{\tilde{P}}

\newcommand{\may}{\tilde{h}_{ij}(k)}

\newcommand{\tG}{\tilde{G}}

\newcommand{\bpsi}{\mathbf{\Psi}}

\newcommand{\ce}{_{\rm c}}
\newcommand{\G}{_{\rm G}}

\newcommand{\h}{_{\rm n}}
\newcommand{\lo}{_{\rm l}}

\newcommand{\n}{_{1\rm p}}

\newcommand{\s}{_{\rm s}}

\newcommand{\hn}{\hat{n}}
\newcommand{\hc}{\hat{\rho}_{\rm c}}
\newcommand{\hh}{\hat{\rho}_{\rm n}}

\begin{document}

\title{Statistical Physics of the Two-Dimensional Coulomb Liquid with Ionic Hard-Core Size}

\author{Sahin Buyukdagli}
\address{Department of Physics, Bilkent University, Ankara 06800, Turkey}

\begin{abstract} 

A self-consistent theory of bulk electrolytes incorporating electrostatic and hard-core interactions on an equal level is applied to the two-dimensional Coulomb liquid with finite ion size. The ionic pair distributions, the structure factors, and the thermodynamic functions of the formalism are compared with extensive Monte-Carlo simulation results from the literature. At moderate salt densities, our computational approach can accurately describe the thermodynamics of two-dimensional solutions from weak to intermediate coupling strengths. The improved accuracy of the present theory with respect to continuum approaches stems mainly from its ability to account for the non-uniform screening of electrostatic interactions associated with the impenetrability of the charged hard disks by their ionic atmosphere. At low salt densities, the validity domain of our self-consistent framework underestimating the extent of ionic cluster formation drops below the critical coupling domain where the conductor-insulator transition of two-dimensional charged hard disks occurs. This indicates that approaching the low-temperature dielectric phase via the present formalism will require the extension of the underlying self-consistent approximation at least up to the next cumulant order.

\end{abstract}

\pacs{05.20.Jj,82.45.Gj,82.35.Rs}
\date{\today}
\maketitle   

\section{Introduction}

Ionic solutions are at the heart of the electrochemical processes fueling carbon-based life on Earth. These complex liquids are particularly difficult to characterize even under the assumption of thermodynamic equilibrium conditions. The complications associated with the thermodynamic formulation of electrolytes originate from the non-integrability of the partition function embodying the long-range Coulomb interactions, and the non-trivial treatment of the ionic hard-core (HC) size incorporated via a discontinuous potential function. 

Due to these technical challenges, any attempt to characterize the thermodynamics of charged fluids has to rely on approximations. For example, the Debye-H\"{u}ckel (DH) formalism introduced as the first quantitatively reliable theory of ionic solutions is based on two major approximations~\cite{DH}. Namely, the theory neglects explicit charge correlations and also HC interactions. As a result, its predictions are limited to dilute monovalent aqueous electrolytes governed by weakly coupled charge interactions.

The substantial part of the parameter regime located beyond the validity domain of the DH approach can be reached by Monte-Carlo (MC) simulations~\cite{Valleau,Svensson,NetzMC,NetzMC2} and integral equation theories~\cite{Hansen,Blum,Henderson,Boda} capable of incorporating unambiguously the ionic HC interactions. While numerical simulations have the obvious advantage of predicting the thermodynamic functions exactly, they are also highly time consuming. Moreover, the explicit solution of the formally exact Ornstein-Zernike identity involves an approximate closure relation whose formulation is not guided by a systematic approach. These points indicate the necessity to develop analytically tractable frameworks enabling the characterization of ionic liquids via transparent and controlled approximations.

The field theoretic description of Coulomb interactions formulated along these lines was introduced by Kholodenko and  Beyerlein for bulk electrolytes~\cite{Kholodenko1,Kholodenko2}. This powerful theoretical framework has been subsequently generalized to nanoconfined electrolytes via weak-coupling (WC)-level loop expansion approaches~\cite{PodWKB,NetzLoop} and self-consistent (SC) formalisms~\cite{netzvar,HatloVar0,BuyukSC}, and strong-coupling-level virial expansion techniques~\cite{NetzSC}. With the aim to access the intermediate coupling regime, Santangelo introduced a splitting technique enabling the asymmetric treatment of the short- and long range ion interactions~\cite{Santangelo}. Subsequently,  Hatlo and Lue upgraded Santangelo's approach via the introduction of a variational splitting scheme enabling the SC interpolation of the theory between the weak- and strong-coupling regimes of ion-macromolecule interactions~\cite{HatloSoft,HatloEpl,LueSoft}. 

The predictions of these continuum theories neglecting ion size are limited to submolar salt densities. In order to extend the field-theoretic formulation of electrostatic systems into the molar density regime where ionic HC size plays a dominant role, we have recently developed a cumulant theory of bulk solutions explicitly incorporating the ionic HC interactions~\cite{Buyuk2024} beyond the dilute salt regime covered by the virial approach formulated by Netz et al.~\cite{NetzHC1,NetzHC2}. Subsequently, we improved the accuracy of this theory by introducing a generalized splitting scheme that allows to avoid the WC-level treatment of the strongly coupled short-range HC and electrostatic interactions~\cite{Soft2024,Buyuk2026}. Via systematic comparison with MC results, we showed that the corresponding self-consistent DH (SCDH) theory can accurately predict the thermodynamics of 3D electrolytes up to molar salt concentrations and also over a broad range of ionic HC sizes~\cite{Buyuk2026}.

While the aforementioned studies of three-dimensional (3D) electrolytes have  been essentially motivated by the need to understand the electrostatically driven mechanisms omnipresent in chemical and biological processes, the interest in two-dimensional (2D) solutions has been mainly triggered by the discovery of a new type of critical behaviour in 2D systems by Berezinskii~\cite{Berez1,Berez2} and Kosterlitz-Thouless~\cite{Koster} in the early 1970s. This breakthrough led to the 2016 Nobel Prize in Physics. 

The Berezinskii-Kosterlitz-Thouless (BKT) transition originally discovered in the XY model is set by the formation of vortex-antivortex pairs at low temperatures. The BKT criticality emerges as well in 2D Coulomb liquids whose charges interact via the logarithmic potential $U=-\Gamma \ln (r/a)$, where $a$ is the ionic radius. In the infinite dilution limit, this transition takes place at the coupling strength $\Gamma=4$ between a conducting phase associated with free charges and a dielectric phase characterized by bound pairs of opposite charges.

One of the earliest evaluations of the internal energy of 2D Coulombic liquids has been carried out via the random phase approximation by Seyler~\cite{Seyler}. Subsequently, Deutsch used a low-temperature binary approximation to calculate the thermodynamic functions of the 2D plasma without repulsive interactions, and showed that the system exhibits an unstability at the coupling constant $\Gamma=2$~\cite{Deutsch}. Additionally, Swendsen ran MC simulations of the 2D liquid on a lattice, and also carried out the analytical evaluation of the thermodynamic functions via the high temperature expansion and the low-temperature independent pair approximation. The confluence of these distinct approaches ascertained the location of the BKT transition at the coupling $\Gamma\ce=4$~\cite{Swendsen}.

The first MC simulations of the 2D Coulomb liquid with actual HC radius has been carried out by Caiolloi and Levesque~\cite{Cailloi}. This study showed that the critical coupling constant increases with the ion density and size. Consequently, at finite salt densities, the BKT transition takes place in the coupling regime $\Gamma\ce>4$. The BKT criticality of the 2D Coulomb liquid has been also analyzed via the low fugacity expansion enabling the access to the dielectric phase near the zero-density regime~\cite{Alastuey,Alastuey2,Kalinay}. Subsequently, Samaj evaluated exactly the thermodynamic functions of the 2D Coulomb liquid within the stability regime $\Gamma<2$ of the vanishing HC size limit~\cite{Samaj2}. Tellez identified as well the like-charge attraction conditions in 2D electrolytes~\cite{Tellez}. Finally, the thermodynamics of the 2D Coulomb liquid with finite HC size has been thoroughly characterized from low to intermediate ion densities via recent numerical simulations~\cite{Lomba,Aupic}.

With the aim to analyze the thermodynamics of 2D charged liquids at finite densities $\rho a^2>0$, in the present work, the SCDH formalism originally introduced for 3D Coulomb fluids~\cite{Soft2024,Buyuk2026} is applied to 2D solutions. Our article is organized as follows. In Sec.~\ref{Mod}, we explain the formulation of the SCDH approach for 2D electrolytes. For the sake of reproducibility of the results, the final identities of the formalism solely enabling the computation of the thermodynamic functions are summarized in Sec.~\ref{resdim}. Sec.~\ref{res} is devoted to the comparison of the thermodynamic functions obtained from our approach with numerical simulations. Therein, the radial pair distributions, the structure factors, the internal energy, and the specific heat of the 2D electrolyte computed within the SCDH formalism are compared with a large amount of MC data from the literature~\cite{Aupic,Lomba,Cailloi}. This comparative analysis enables us to identify accurately the validity domain of the present approach in terms of the electrostatic coupling parameter $\Gamma$ and the dimensionless salt density $\rho a^2$. Finally, the main progress and the limitations of the present formalism, and potential improvements and extensions of the underlying theoretical framework are elaborated in Conclusions.

\section{2D SCDH formalism}

In this section, we introduce a modified version of the SCDH formalism previously developed in Refs.~\cite{Soft2024,Buyuk2026} for 3D solutions. These modifications consist in (i) replacing the filter operator of quartic order used in Refs.~\cite{Soft2024,Buyuk2026} by an eight order operator (see Eqs.~(\ref{eq18})-(\ref{eq19}) below), (ii) limiting the long-range correlation function to gaussian level, and (iii) reformulating the theoretical framework for 2D space.

\label{Mod}

\subsection{2D Liquid Model and Partition Function}

The bulk liquid is composed of $p$ ion species. Each ion of the species $i$ with valency $q_i$, concentration $n_i$, and  fugacity $\lambda_i$ is placed at the center of a HC disk with diameter $a$. In the liquid, two ions of positions $\br$ and $\br'$ and separation distance $r\s=||\br-\br'||$ interact via the HC potential $v\h(r\s)$ defined by the identity
\be
\label{eq1}
e^{-v\h(r\s)}=\theta(r\s-a)
\ee
involving the Heaviside step function $\theta(x)$~\cite{math}, and the 2D Coulomb potential
\be\label{eq2}
v\ce(r\s)=-\Gamma\ln\left(\frac{r\s}{a}\right)
\ee
solving the kernel equation
\be
\label{eq3}
\nabla^2 v\ce(\br,\br')=-2\pi\Gamma\delta^2(\br-\br').
\ee
Eqs.~(\ref{eq2})-(\ref{eq3}) include the electrostatic coupling parameter $\Gamma=\beta e^2/(2\pi\e_0\e_wa)$ with the inverse thermal energy $\beta=1/(k_{\rm B}T)$, where $k_{\rm B}$ is the Boltzmann constant, $T$ stands for the ambient temperature, $e$ is the electron charge, and $\e_0$ and $\e_w$ are the dielectric permittivity of vacuum and the relative dielectric constant of water, respectively.

Combining now Eq.~(\ref{eq3})  with the definition of the inverse $G^{-1}(\br,\br')$ of the general Green's function $G(\br,\br')$,
\be
\label{eq3II}
\int\mathrm{d}^2\br_1G^{-1}(\br,\br_1)G(\br_1,\br')=\delta^2(\br-\br'),
\ee
the inverse of the Coulomb potential~(\ref{eq2}) required for the formulation of the splitting scheme introduced below follows as
\be
\label{eq3III}
v\ce^{-1}(\br,\br')=-\frac{1}{2\pi\Gamma}\nabla^2\delta^2(\br-\br').
\ee

The Grand-Canonical (GC) partition function of the liquid is defined as the trace over the fluctuating ion numbers $N_i$ and positions $\br_{ij}$,
\be\label{eq4}
Z\G=\prod_{i=1}^p\sum_{N_i=0}^\infty\frac{\lambda_i^{N_i}}{N_i!}\prod_{j=1}^p\prod_{k=1}^{N_j}\int\mathrm{d}^2\br_{jk}\;e^{-\beta(E\ce+E\h+E\n-E\s)}.
\ee
In Eq.~(\ref{eq4}), the pairwise energy components 
\be
\label{eq5}
\beta E_\gamma=\frac{1}{2}\int\mathrm{d}^2\br\mathrm{d}^2\br'\hat{\rho}_\gamma(\br)v_\gamma(\br,\br')\hat{\rho}_\gamma(\br')
\ee
incorporating the Coulombic charge coupling ($\gamma={\rm c}$) and HC interactions ($\gamma={\rm n}$) have been expressed in terms of the following ion number and charge density operators,
\be
\label{eq8}
\hh(\br)=\sum_{i=1}^p\hn_i(\br);\hspace{5mm}\hc(\br)=\sum_{i=1}^pq_i\hn_i(\br),
\ee
where the number density operator of the ionic species $i$ is defined as
\be
\label{eq9}
\hn_i(\br)=\sum_{j=1}^{N_i}\delta^2(\br-\br_{ij}).
\ee 
Moreover, the Boltzmann distribution in Eq.~(\ref{eq4}) contains the one-body potential energy 
\be
\label{eq6}
\beta E\n=\sum_{i=1}^p\sum_{j=1}^{N_i}w_{i}(\br_{ij})=\sum_{i=1}^p\int\mathrm{d}^2\br\;w_i(\br)\hn_i(\br)
\ee
to be used for the derivation of the average ion densities, and the total self-energy 
\be
\label{eq7}
\beta E\s=\sum_{i=1}^p\sum_{j=1}^{N_i}\epsilon_i
\ee
to be subtracted from the interaction energy, where $\epsilon_i=\left[q_i^2v\ce(0)+v\h(0)\right]/2$ is the self energy of a single charge.

Following the splitting technique of Refs.~\cite{Santangelo,HatloSoft,HatloEpl,LueSoft,Soft2024,Buyuk2026} developed for 3D liquids, we separate now the Coulomb potential~(\ref{eq2}) into two parts associated with distinct spatial ranges,
\be\label{eq10}
v\ce(\br,\br')=v\s(\br,\br')+v\lo(\br,\br'),
\ee
where the functional form of the potentials $v_\alpha(\br,\br')$ incorporating the short-range ($\alpha={\rm s}$) and long-range ($\alpha={\rm l}$) ion interactions will be specified below. In the Boltzmann distribution of Eq.~(\ref{eq4}), the splitting~(\ref{eq10}) generates two types of pairwise electrostatic interactions. Thus, in Eq.~(\ref{eq4}), we introduce two separate Hubbard-Stratonovich (HS) transformations of the form
\bea
\label{eq11}
&&e^{-\frac{1}{2}\int\mathrm{d}^2\br\mathrm{d}^2\br'\hc(\br)v_\alpha(\br,\br')\hc(\br')}\\
&&=\int\frac{\mathcal{D}\psi_\alpha}{\sqrt{{\rm det}\left[v_\alpha\right]}}\;e^{-\frac{1}{2}\int\mathrm{d}^2\br\mathrm{d}^2\br'\psi_\alpha(\br)v^{-1}_\alpha(\br,\br')\psi_\alpha(\br')}\nonumber\\
&&\hspace{2.25cm}\times e^{i\int\mathrm{d}^2\br\;\hc(\br)\psi_\alpha(\br)}\nonumber
\eea
for the corresponding charge interactions ($\alpha=\{{\rm s,l}\}$), and the additional HS transformation
\bea
\label{eq12}
&&e^{-\frac{1}{2}\int\mathrm{d}^2\br\mathrm{d}^2\br'\hh(\br)v\h(\br,\br')\hh(\br')}\\
&&=\int\frac{\mathcal{D}\psi\h}{\sqrt{{\rm det}\left[v\h\right]}}\;e^{-\frac{1}{2}\int\mathrm{d}^2\br\mathrm{d}^2\br'\psi\h(\br)v^{-1}\h(\br,\br')\psi\h(\br')}\nonumber\\
&&\hspace{2.25cm}\times e^{i\int\mathrm{d}^2\br\;\hh(\br)\psi\h(\br)}\nonumber
\eea
for the HC interactions, where the functions $\psi_\alpha(\br)$ are the fluctuating potentials associated with the HC ($\alpha={\rm n}$) and electrostatic coupling ($\alpha=\{{\rm s,l}\}$). These transformations allow us to evaluate the geometric sums in the GC partition function~(\ref{eq4}) and to recast the latter in the form of the following functional integral
\be\label{eq14}
Z\G=\int\frac{\mathcal{D}\bpsi}{\sqrt{{\rm det}\left[v\h v\s v\lo\right]}}e^{-\beta H[\bpsi]}
\ee
including the Hamiltonian functional
\bea
\label{eq15}
\beta H[\bpsi]&=&\sum_{\alpha=\{{\rm n,s,l}\}}\int\frac{\mathrm{d}^2\br\mathrm{d}^2\br'}{2}\psi_\alpha(\br)v_\alpha^{-1}(\br,\br')\psi_\alpha(\br')\nonumber\\
&&-\sum_{i=1}^p\lambda_i\int\mathrm{d}^2\br \;\hk_i(\br).
\eea
In Eqs.~(\ref{eq14})-(\ref{eq15}), we introduced the shorthand  vector notations for the fluctuating potentials $\bpsi=(\psi\h,\psi\s,\psi\lo)$ and the functional integration measure $\mathcal{D}\bpsi=\mathcal{D}\psi\h\mathcal{D}\psi\s\mathcal{D}\psi\lo$, and the dimensionless fluctuating ion density 
\be\label{eq16}
\hk_i(\br)=e^{\epsilon_i-w_i(\br)+i\psi\h(\br)+iq_i\left[\psi\s(\br)+\psi\lo(\br)\right]}.
\ee

\subsection{Varitational Splitting Scheme}

We introduce here the variational splitting scheme that will enable the asymmetric treatment of the short- and long-range ion interactions. In the present article, the 2D Fourier-Transform (FT) of the general function $f(\br)$ and its inverse FT are defined as $\tilde{f}(k)=\int\mathrm{d}^2\br f(\br)e^{-i\bk\cdot\br}$ and $f(\br)=(2\pi)^{-2}\int\mathrm{d}^2\bk \tilde{f}(k)e^{i\bk\cdot\br}$, respectively. Expressing now the kernel identity~(\ref{eq3}) in reciprocal space, the FT of the 2D Coulomb potential~(\ref{eq2}) follows as
\be
\label{eq17}
\tv\ce(k)=\frac{2\pi\Gamma}{k^2}.
\ee

Our specific choice of the long-range potential in Eq.~(\ref{eq10}) can be expressed in terms of the inverse Coulomb operator~(\ref{eq3III}) as
\be
\label{eq18}
v\lo^{-1}(\br,\br')=\hat{P}v\ce^{-1}(\br,\br'),
\ee
where the eight order differential operator
\be
\label{eq19}
\hat{P}=1-\sigma^2\nabla^2+\sigma^4\nabla^4-\sigma^6\nabla^6+\sigma^8\nabla^8
\ee
including the characteristic splitting length $\sigma$ to be determined variationally filters out the short wavelengths. The choice of an eight order differential operator that differs from the quartic order operator of earlier works~\cite{Santangelo,HatloSoft,HatloEpl,LueSoft,Soft2024,Buyuk2026} is motivated by our observation of an improved agreement with MC simulation results by the increase of the highest differential order in Eq.~(\ref{eq19})~\cite{rem1}.

Inverting now Eq.~(\ref{eq18}) in Fourier space, and using the constraint~(\ref{eq10}) and Eq.~(\ref{eq17}), the long- and short-range pairwise potentials follow as the 2D Fourier integrals
\bea
\label{eq20}
v\lo(r)&=&\Gamma\int_0^\infty\frac{\mathrm{d}k}{k}\frac{J_0(kr)}{\tP(k)};\\
\label{eq20II}
v\s(r)&=&\Gamma\int_0^\infty\frac{\mathrm{d}k}{k}\frac{\tP(k)-1}{\tP(k)}J_0(kr),
\eea
where $J_n(x)$ is the Bessel function of the first kind~\cite{math}, and the FT of the operator~(\ref{eq19}) reads
\be\label{eq21}
\tP(k)=1+\sigma^2k^2+\sigma^4k^4+\sigma^6k^6+\sigma^8k^8.
\ee

Our derivation of the variational identity satisfied by the splitting parameter $\sigma$ will be based on the invariance of the partition function~(\ref{eq4}) and the grand potential $\Omega\G=-k_{\rm B}T\ln Z\G$ on this characteristic length. Thus, evaluating the equation $\partial_\sigma\Omega\G=0$ with the functional integral form~(\ref{eq14}) of the partition function, one obtains
\be
\label{eq22}
\sum_{\alpha=\{{\rm s,l}\}}\int\mathrm{d}^2\br\mathrm{d}^2\br'\left[G_\alpha(\br,\br')-v_\alpha(\br,\br')\right]\partial_\sigma v^{-1}_\gamma(\br,\br')=0,
\ee
where the two-point correlation function (2PCF) 
\be
\label{eq23}
G_\alpha(\br,\br')=\lan\psi_\alpha(\br)\psi_\alpha(\br')\ran
\ee
involves the field-theoretic average defined for the general functional $F[\bpsi]$ as
\be
\label{eq24}
\lan F[\bpsi]\ran=\frac{1}{Z\G}\int\frac{\mathcal{D}\bpsi}{\sqrt{{\rm det}\left[v\h v\s v\lo\right]}} e^{-\beta H[\bpsi]}F[\bpsi].
\ee
Finally, accounting for the translational invariance in the bulk liquid implying $v_\alpha(\br,\br')=v_\alpha(\br-\br')$ and $G_\alpha(\br,\br')=G_\alpha(\br-\br')$, one can express Eq.~(\ref{eq22}) in Fourier space as
\be
\label{eq25}
\sum_{\alpha=\{{\rm s,l}\}}\int_0^\infty\mathrm{d}kk\left[\tG_\alpha(k)-\tv_\alpha(k)\right]\tv^{-2}_\alpha(k)\partial_\sigma\tv_\alpha(k)=0,
\ee
where the FT of the long- and short-range potential components in Eqs.~(\ref{eq20})-(\ref{eq20II}) read
\be
\label{eq26}
\tv\lo(k)=\frac{2\pi\Gamma}{k^2\tP(k)};\hspace{5mm}\tv\s(k)=\frac{2\pi\Gamma}{k^2}\frac{\tP(k)-1}{\tP(k)}.
\ee

At this point, the formally exact identity~(\ref{eq25}) is satisfied by any value of the splitting parameter $\sigma$. Below, Eq.~(\ref{eq25}) will become the variational identity solved by the specific value of $\sigma$ satisfying the stationary state condition of the grand potential ($\partial_\sigma\Omega\G=0$) upon the introduction of the approximation scheme enabling the explicit evaluation of the 2PCF~(\ref{eq23}).

\subsection{Ion Density and Pair Distribution Functions}

We derive now the relationship between the ionic fugacity and concentration. The concentration of the ionic species $i$ corresponds to the GC average of the density operator in Eq.~(\ref{eq9}), i.e. $n_i=\lan\hn_i(\br)\ran\G$. According to Eqs.~(\ref{eq4}) and~(\ref{eq6}), this average can be obtained from the relation $n_i=-Z\G^{-1}\delta Z\G/\delta w_i(\br)$. Evaluating the latter identity with the functional integral representation~(\ref{eq14}) of the partition function, one obtains
\be
\label{eq27}
n_i=\lambda_i\lan \hk_i(\br)\ran.
\ee

The computation of the thermodynamic variables of the charged liquid requires the knowledge of the pair distribution function associated with two ions of the species $i$ and $j$. The latter is defined as the GC average
\be
\label{eq28}
n_in_jg_{ij}(\br,\br')=\lan\hn_i(\br)\hn_j(\br')\ran\G-\delta_{ij}\lan\hn_i(\br)\ran\G\delta^2(\br-\br'),
\ee
where the negative term including the Kronecker delta symbol $\delta_{ij}$ subtracts the self-interactions. Via Eqs.~(\ref{eq4}), (\ref{eq6}), and~(\ref{eq27}), Eq.~(\ref{eq28}) can be now expressed as 
\be
\label{eq29}
n_in_jg_{ij}(\br,\br')=\frac{1}{Z\G}\frac{\delta^2 Z\G}{\delta w_i(\br)\delta w_j(\br')}-n_i\delta_{ij}\delta^2(\br-\br').
\ee
Finally, inserting the functional integral form~(\ref{eq14}) of the partition function into Eq.~(\ref{eq29}), the pair distribution function follows as a functional average of the form
\be
\label{eq30}
g_{ij}(\br,\br')=\frac{\lambda_i\lambda_j}{n_in_j}\lan \hk_i(\br)k_j(\br')\ran.
\ee

\subsection{Electroneutrality}

A significant part of the formally exact identities derived in the present article will be based on the Schwinger-Dyson (SD) equation 
\be\label{eq31}
\lan\frac{\delta F[\bpsi]}{\delta\psi_\alpha(\br)}\ran=\lan F[\bpsi]\frac{\delta H[\bpsi]}{\delta\psi_\alpha(\br)}\ran
\ee
relating the functional derivatives of the Hamiltonian~(\ref{eq15}) and the general functional $F[\bpsi]$~\cite{justin}. The derivation of the SD equation~(\ref{eq31}) from the invariance of the functional integral~(\ref{eq24}) under an infinitesimal shift of the potentials $\psi_\alpha(\br)$ can be found in Refs.~\cite{Soft2024,Buyuk2026}.

In order to derive the global electroneutrality condition, in Eq.~(\ref{eq31}), we first set $F[\bpsi]=1$ and $\alpha={\rm l}$. This yields $\lan\delta H[\bpsi]/\delta\psi\lo(\br)\ran=0$. Plugging into this relation the Hamiltonian functional~(\ref{eq15}), using Eq.~(\ref{eq27}), and passing to Fourier space, one obtains the identity $\tv\lo^{-1}(0)\bar{\psi}\lo=i\sum_in_iq_i$, where we accounted for the uniformity of the average potential $\bar{\psi}\lo=\lan\psi\lo(\br)\ran$ originating from the translational symmetry in the bulk liquid. Noting now the infrared (IR) cancellation of the inverse of the Fourier-transformed long-range potential in Eq.~(\ref{eq26}), i.e. $\tv\lo^{-1}(k\to0)=0$, one finally obtains the global electroneutrality constraint
\be
\label{eq32}
\sum_{i=1}^pn_iq_i=0.
\ee

\subsection{Evaluation of the Functional Averages}

The computation of the thermodynamic variables of the 2D liquid requires the explicit evaluation of the functional averages of the form~(\ref{eq24}) involved in the formally exact identities that have so far been derived. In this part, we introduce a simplified version of the SCDH formalism~\cite{Soft2024,Buyuk2026} that will enable the approximate calculation of these thermodynamic functions. In order to switch to a compact notation, from now on, we will omit the potential dependence of the functionals. 

\subsubsection{Mixed Expansion Scheme}

Our approximation scheme that will enable the asymmetric treatment of the short- and long-range ion interactions is based on the following splitting of the Hamiltonian~(\ref{eq15}),
\be
\label{eq38}
H=H_0+t\delta H.
\ee
In Eq.~(\ref{eq38}), $H_0$ is the reference gaussian Hamiltonian that will be treated exactly. Then, the additional component $\delta H$ accounting for the non-linear potential fluctuations beyond the gaussian-level will be incorporated approximately via a perturbative expansion in terms of the parameter $t$. The corresponding parameter of unit magnitude ($t=1$) will enable us to keep track of the expansion order.

The SCDH formalism is based on the following definition of the gaussian-level reference Hamiltonian,
\bea
\label{eq41}
\beta H_0&=&\frac{1}{2}\sum_{\alpha=\{{\rm n,s}\}}\int\mathrm{d}^2\br\mathrm{d}^2\br'\psi_\alpha(\br)v_\alpha^{-1}(\br,\br')\psi_\alpha(\br')\nonumber\\
&&+\frac{1}{2}\int\mathrm{d}^2\br\mathrm{d}^2\br'\psi\lo(\br)G\lo^{-1}(\br,\br')\psi\lo(\br').
\eea
Within the approximation scheme defined by the specific choice in Eq.~(\ref{eq41}), the short-range HC and electrostatic interactions ($\alpha=\{{\rm n,s}\}$) are included via the bare pairwise interactions $v_\alpha(\br,\br')$, and the long-range correlations are incorporated by the unknown kernel $G\lo(\br,\br')$ to be determined from the SC solution of the SD Eqs.~(\ref{eq31}). This mixed approximation scheme corresponding to the virial treatment of the strongly coupled short-range interactions will precisely enable to avoid their WC-level gaussian treatment applied only to the screened long-range interactions. We finally note that via Eq.~(\ref{eq15}), the non-linear Hamiltonian component in Eq.~(\ref{eq38}) follows as
\bea
\label{eq42}
\beta\delta H&=&\frac{1}{2}\int\mathrm{d}^2\br\mathrm{d}^2\br'\psi\lo(\br)\left[v\lo^{-1}-G\lo^{-1}\right]_{\br,\br'}\psi\lo(\br')\nonumber\\
&&-\sum_{i=1}^p\lambda_i\int\mathrm{d}^2\br\;\hk_i(\br).
\eea

Substituting now the identity~(\ref{eq38}) into Eq.~(\ref{eq24}), and Taylor expanding the result at the order $O(t)$, the field-theoretic average of the functional $F$ follows as
\be
\label{eq39}
\lan F\ran=\lan F\ran_0-t\left[\lan\beta\delta HF\ran_0-\lan\beta\delta H\ran_0\lan F\ran_0\right]+O(t^2),
\ee
where the gaussian-level functional average and partition function read
\be\label{eq40}
\lan F\ran_0=\frac{1}{Z_0}\int\mathcal{D}\bpsi e^{-\beta H_0}F;\hspace{5mm}Z_0=\int\mathcal{D}\bpsi e^{-\beta H_0}.
\ee

\subsubsection{Relating Ionic Fugacity and Concentration}

Here, we carry out the explicit evaluation of the relationship~(\ref{eq27}) between the ion fugacity and concentration. To this aim, we calculate the functional average in Eq.~(\ref{eq27}) according to Eqs.~(\ref{eq39})-(\ref{eq40}) to obtain
\bea
\label{eq43}
n_i&\approx&\Lambda_i+t\Lambda_i\sum_{j=1}^p\Lambda_j\int\mathrm{d}^2\br\;h_{ij}(\br)\\
&&+t\Lambda_i\frac{q_i^2}{2}\int\mathrm{d}^2\br_1\mathrm{d}^2\br_2\left[v\lo^{-1}(\br_1,\br_2)-G\lo^{-1}(\br_1,\br_2)\right]\nonumber\\
&&\hspace{2.7cm}\times G\lo(\br,\br_1)G\lo(\br_2,\br).\nonumber
\eea
In Eq.~(\ref{eq43}), we introduced the rescaled fugacity $\Lambda_i=\lambda_i\;e^{-\frac{q_i^2}{2}\left[G\lo(0)-v\lo(0)\right]}$, and the Mayer function
\be
\label{eq44}
h_{ij}(r)=\theta(r-a)\;e^{-q_iq_j\left[G\lo(r)+v\s(r)\right]}-1.
\ee

In order to invert the identity~(\ref{eq43}), we insert into the latter the formal expansion $\Lambda_i=\Lambda_i^{(0)}+t\Lambda_i^{(1)}+O(t^2)$ and identity the coefficients $\Lambda_i^{(n)}$ of different perturbative orders. At the order $O(t)$, one finally gets the ion fugacity as a function of the ionic concentration $n_i$ as
\bea
\label{eq45}
\Lambda_i&\approx&n_i-tn_i\sum_{j=1}^pn_j\int\mathrm{d}^2\br\;h_{ij}(\br)\\
&&+tn_i\frac{q_i^2}{2}\int\mathrm{d}^2\br_1\mathrm{d}^2\br_2\left[G\lo^{-1}(\br_1,\br_2)-v\lo^{-1}(\br_1,\br_2)\right]\nonumber\\
&&\hspace{2.7cm}\times G\lo(\br,\br_1)G\lo(\br_2,\br).\nonumber
\eea

\subsubsection{Calculation of the variational identity~(\ref{eq25})}

The explicit evaluation of the variational identity~(\ref{eq25}) requires the calculation of the correlators $G_\alpha(\br,\br')$. To this aim, in Eq.~(\ref{eq31}), we set $F[\bpsi]=\psi_\alpha(\br')$. This yields
\bea
\label{eq46I}
&&\int\mathrm{d}^2\br_1v_\alpha^{-1}(\br,\br_1)\lan\psi_\alpha(\br_1)\psi_\alpha(\br')\ran\\
&&-i\sum_{i=1}^p\lambda_iq_i\lan\hk_i(\br)\psi_\alpha(\br')\ran=0.\nonumber
\eea
Upon the inversion of Eq.~(\ref{eq46I}) via Eq.~(\ref{eq3II}), one obtains
\bea
\label{eq46}
&&G_\alpha(\br,\br')-v_\alpha(\br,\br')\\
&&=i\sum_{i=1}^p\lambda_iq_i\int\mathrm{d}^2\br_1v_\alpha(\br,\br_1)\lan\hk_i(\br_1)\psi_\alpha(\br')\ran\nonumber
\eea

In Eq.~(\ref{eq46}), we first set $\alpha={\rm s}$, and evaluate the resulting functional average according to Eq.~(\ref{eq39}). Plugging the ion fugacity~(\ref{eq45}), and Taylor-expanding the result at the order $O(t)$, after lengthy algebra, one gets
\bea
\label{eq47}
G\s(\br,\br')&\approx&v\s(\br,\br')-\sum_{i=1}^pn_iq_i^2\int\mathrm{d}^2\br_1v\s(\br,\br_1)v\s(\br_1,\br')\nonumber\\
&&-t\sum_{i,j}n_in_jq_iq_j\int\mathrm{d}^2\br_1\mathrm{d}^2\br_2\;h_{ij}(\br_1,\br_2)\\
&&\hspace{3.7cm}\times v\s(\br,\br_1)v\s(\br_2,\br').\nonumber
\eea
At this point, we introduce the simplified feature of the present formalism with respect to our earlier works in Refs.~\cite{Soft2024,Buyuk2026}. This simplification consists in treating the long-range interactions at the purely gaussian-level. Thus, setting in Eq.~(\ref{eq46}) $\alpha={\rm l}$, computing the corresponding functional average via Eq.~(\ref{eq39}), inserting the ion fugacity~(\ref{eq45}) into the resulting expression, and expanding the result at the order $O(t^0)$, one gets~\cite{rem2}
\be
\label{eq48}
G\lo(\br,\br')\approx v\lo(\br,\br')-\sum_{i=1}^pn_iq_i^2\int\mathrm{d}^2\br_1v\lo(\br,\br_1)G\lo(\br_1,\br').
\ee

The Fourier transformation of Eqs.~(\ref{eq47})-(\ref{eq48}) now yields
\bea
\label{eq49}
\tG\s(k)-\tv\s(k)&=&-\sum_{i=1}^pn_iq_i^2\tv^2\s(k)\\
&&-t\sum_{i,j}n_in_jq_iq_j\tv\s^2(k)\may+O(t);\nonumber\\
\label{eq50}
\tG\lo(k)-\tv\lo(k)&=&-\sum_{i=1}^pn_iq_i^2\tv\lo(k)\tG\lo(k)+O(t^0),
\eea
where the FT of the Mayer function~(\ref{eq44}) reads
\be
\label{eq50II}
\may=-2\pi\frac{a}{k}{\rm J}_1(ka)+2\pi\int_a^\infty\mathrm{d}rr{\rm J}_0(kr)h_{ij}(r).
\ee
From Eq.~(\ref{eq50}), one obtains the long-range kernel in reciprocal and real spaces required for the evaluation of the Mayer function~(\ref{eq44}) and its FT~(\ref{eq50II}) as
\be
\label{eq52}
\hspace{-2mm}\tG\lo(k)=\frac{2\pi\Gamma}{k^2\tP(k)+\kappa_0^2};\hspace{1mm}G\lo(r)=\Gamma\int_0^\infty\frac{\mathrm{d}kk\;{\rm J}_0(kr)}{k^2\tP(k)+\kappa_0^2},
\ee
with the DH screening parameter
\be\label{eq53}
\kappa_0^2=2\pi\Gamma\sum_{i=1}^pn_iq_i^2.
\ee
Finally, plugging Eqs.~(\ref{eq49})-(\ref{eq50}) into the identity~(\ref{eq25}), the variational equation satisfied by the splitting parameter $\sigma$ follows in the compact form
\be\label{eq51}
\sum_{i,j}n_in_jq_iq_j\int_0^\infty\mathrm{d}kk\hspace{-.5mm}\left\{\may+q_iq_j\tG\lo(k)\right\}\partial_\sigma\tv\lo(k)=0.
\ee

\subsubsection{Derivation of the total correlation function}

In order to calculate the total correlation function
\be
\label{eq34II}
H_{ij}(\br,\br')=g_{ij}(\br,\br')-1
\ee
required for the evaluation of the radial distributions and the thermodynamic quantitites, we evaluate the functional average in the pair distribution function~(\ref{eq30}) according to Eqs.~(\ref{eq39})-(\ref{eq40}), replace the ion fugacity by the concentration via Eq.~(\ref{eq45}), and expand the result up to the perturbative order $O(t)$. This yields
\be
\label{eq54}
H_{ij}(\br-\br')\approx h_{ij}(\br-\br')+t\left[h_{ij}(\br-\br')+1\right]T_{ij}(\br-\br'),
\ee
where we introduced the auxiliary function
\bea
\label{eq55}
T_{ij}(\br-\br')&=&\sum_{n=1}^p\int\mathrm{d}^2\br_1\left\{h_{in}(\br-\br_1)h_{nj}(\br_1-\br')\right.\\
&&\left.\hspace{1.5cm}-q_iq_jq_n^2G\lo(\br-\br_1)G\lo(\br_1-\br')\right\}.\nonumber
\eea
Passing to Fourier space, the auxiliary function~(\ref{eq55}) can be now expressed in terms of the Fourier-transformed functions in Eqs.~(\ref{eq50II})-(\ref{eq52}) as
\be
\label{eq56}
T_{ij}(r)\hspace{-1mm}=\hspace{-1mm}\sum_{n=1}^p\int_0^\infty\hspace{-1mm}\frac{\mathrm{d}kk}{2\pi^2}{\rm J}_0(kr)\hspace{-.7mm}\left\{\tilde{h}_{in}(k)\tilde{h}_{nj}(k)-q_iq_jq_n^2\tG\lo^2(k)\right\}.
\ee
\begin{figure*}
\includegraphics[width=.8\linewidth]{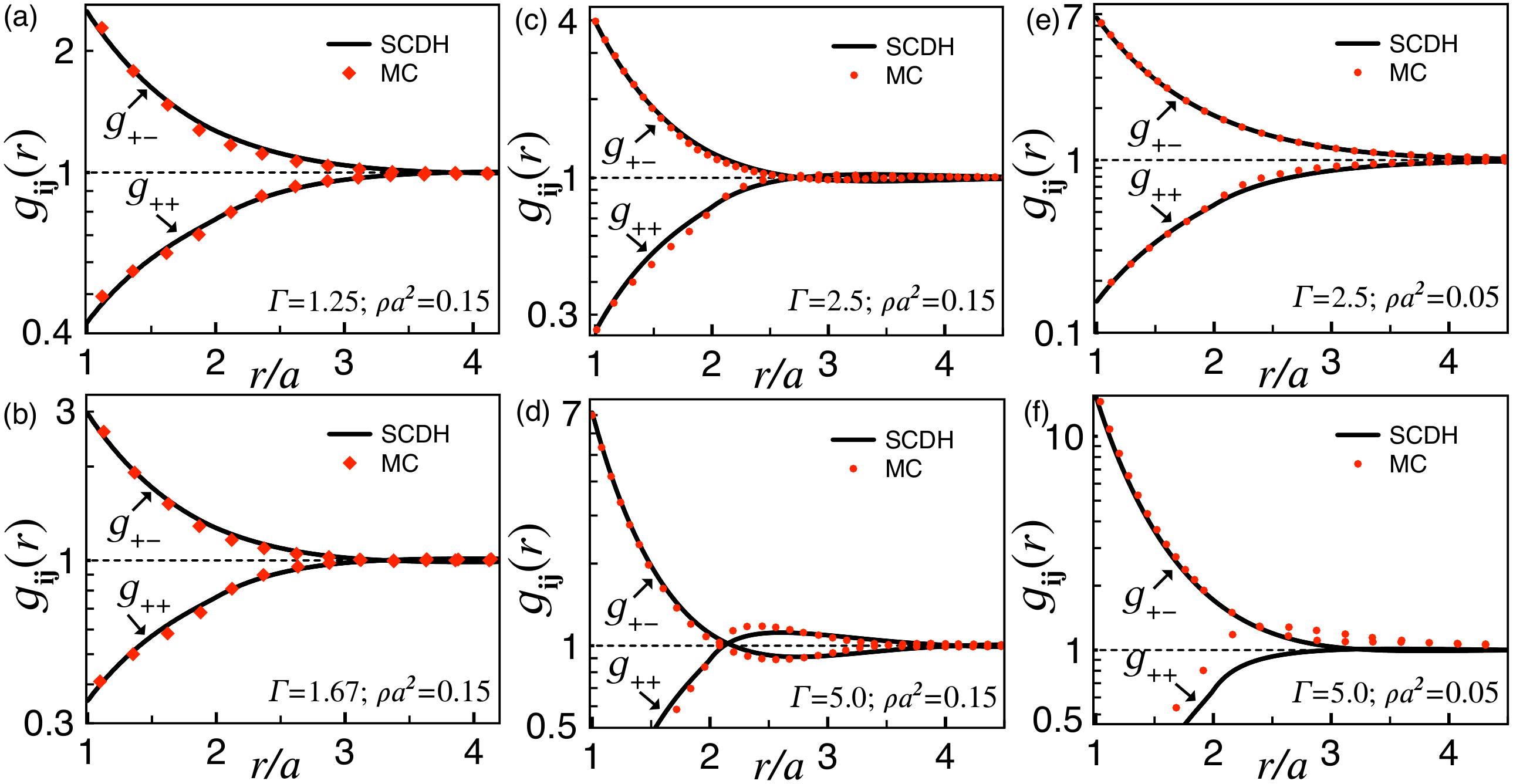}
\caption{(Color online) Pair distribution functions. Symbols: MC data from Figs. 2 and 4 of Ref.~\cite{Aupic} (diamonds) and Fig. 4 of Ref.~\cite{Lomba} (disks). Solid curves: SCDH prediction from Eqs.~(\ref{eq34II}) and~(\ref{eq64}). The values of the electrostatic coupling parameter and the reduced density are indicated in the legends.}
\label{fig1}
\end{figure*}

\subsection{Dimensionless Equations}
\label{resdim}

From now on,  we consider a $1:1$ electrolyte. Thus, we set $q_\pm=\pm1$ and $n_+=n_-$. For this symmetric solution, we recast here the key identities of the formalism required for the computation of the thermodynamic functions in dimensionless form. For the sake of reproducibility of the results, below, this scaling is carried out in the order that should be followed for the numerical implementation of these equations. To this aim, we introduce the dimensionless Fourier variable $q=ka$ and FT $\bar{f}(q)=\tilde{f}(k)/a^2$, the rescaled distance $u=r/a$ and splitting parameter $\bar\sigma=\sigma/a$, and the total ion concentration $\rho=2n_i$. 

Within this notation, the potentials in Eqs.~(\ref{eq20II})  and~(\ref{eq52}) become the dimensionless Fourier integrals
\bea\label{eq57}
v\s(u)&=&\Gamma\int_0^\infty\frac{\mathrm{d}q}{q}\frac{S(q)-1}{S(q)}J_0(qu);\\
\label{eq58}
G\lo(u)&=&\Gamma\int_0^\infty\frac{\mathrm{d}qq}{q^2S(q)+\bar\kappa_0^2}{\rm J}_0(qu),
\eea
with the FT of the filter function~(\ref{eq19}) given by
\be\label{eq59}
S(q)=1+(\bar\sigma q)^2+(\bar\sigma q)^4+(\bar\sigma q)^6+(\bar\sigma q)^8,
\ee
and the rescaled DH parameter 
\be\label{eq59II}
\bar\kappa_0^2=2\pi\Gamma \rho a^2.
\ee
In terms of the potentials~(\ref{eq57})-(\ref{eq58}), the real-space Mayer function~(\ref{eq44}) now reads
\be\label{eq60}
h_{+\pm}(u)=\theta(u-1)\;e^{\mp\left[v\s(u)+G\lo(u)\right]}-1,
\ee
and its FT~(\ref{eq50II}) becomes
\be\label{eq61}
\bar{h}_{ij}(q)=-\frac{2\pi}{q}{\rm J}_1(q)+2\pi\int_1^\infty\mathrm{d}uu\;{\rm J}_0(qu)h_{ij}(u).
\ee
Expressing as well the dimensionless FT of the long-range potentials in Eqs.~(\ref{eq26}) and~(\ref{eq52}) as
\be\label{eq62}
\bar{v}\lo(q)=\frac{2\pi\Gamma}{q^2S(q)};\hspace{5mm}\bar{G}\lo(q)=\frac{2\pi\Gamma}{q^2S(q)+\bar\kappa_0^2},
\ee
the variational Eq.~(\ref{eq51}) can be recast in the rescaled form
\be\label{eq63}
\int_0^\infty\mathrm{d}qq\hspace{-.5mm}\left\{\bar{h}_{++}(q)-\bar{h}_{+-}(q)+2\bar{G}\lo(q)\right\}\partial_{\bar\sigma}\bar{v}\lo(q)=0.
\ee

In the present work, a standard dichotomy algorithm was employed for the solution of the variational identity~(\ref{eq63}). We finally note that within the same dimensionless notation, the correlation function~(\ref{eq54}) becomes
\be\label{eq64}
H_{ij}(u)=h_{ij}(u)+\left[h_{ij}(u)+1\right]T_{ij}(u),
\ee
where the components of the auxiliary functions~(\ref{eq56}) can be expressed as the dimensionless Fourier integrals
\bea\label{eq65}
T_{++}(u)\hspace{-1mm}&=&\hspace{-1mm}\frac{\rho a^2}{2}\hspace{-2mm}\int_0^\infty\hspace{-1mm}\frac{\mathrm{d}qq}{2\pi^2}{\rm J}_0(qu)\hspace{-.5mm}\left[\bar{h}_{++}^2(q)+\bar{h}^2_{+-}(q)-2\bar{G}\lo^2(q)\right];\nonumber\\
&&\\
\label{eq66}
T_{+-}(u)\hspace{-1mm}&=&\hspace{-1mm}\rho a^2\hspace{-2mm}\int_0^\infty\hspace{-1mm}\frac{\mathrm{d}qq}{2\pi^2}{\rm J}_0(qu)\hspace{-.5mm}\left[\bar{h}_{++}(q)\bar{h}_{+-}(q)+\bar{G}\lo^2(q)\right].
\eea

\section{Results}
\label{res}

With the aim to identify the validity regime of the SCDH formalism in 2D, in this part, we compare the ionic pair distribution functions, the structure factors, and the thermodynamic functions of the 2D Coulomb liquid obtained from the present approach with MC simulation results from the literature.

\subsection{Pair distributions and structure factors}
\label{pr}

In Fig.~\ref{fig1}, we compare the ionic pair distributions of the SCDH formalism obtained from Eqs.~(\ref{eq34II}) and~(\ref{eq64}) (solid curves) with the MC simulation results of Refs.~\cite{Lomba,Aupic} (red symbols).  Figs.~\ref{fig1}(a)-(d) indicate that at the intermediate salt concentration $\rho a^2=0.15$, the SCDH formalism can accurately reproduce the opposite- and like-charge pair distributions of the MC simulations in the intermediate coupling regime extending from $\Gamma=1.25$ up to $\Gamma=5.0$. In particular, one sees that at the highest coupling strength considered in Fig.~\ref{fig1}(d), the opposite- and similar-charge pair distributions of the MC simulations exhibit a minimum and a peak, respectively. These structures originating from the formation of ionic pair clusters~\cite{Lomba} are equally taken into account by our approach with reasonable accuracy. 

We investigate now the accuracy of the SCDH approach against the variation of the salt density. In 2D electrolytes, the decrease of the salt density is known to amplify the ionic cluster formation and to drive the system towards the critical BKT line where the insulating phase arises~\cite{Lomba}. The comparison of Figs.~\ref{fig1}(c) and (e) indicates that at the coupling strength $\Gamma=2.5$, the SCDH formalism can accurately take into account the reduction of the salt density from $\rho a^2=0.15$ down to $0.05$. However, Figs.~\ref{fig1}(d) and (f) show that upon the same decrease of the ion concentration at the higher coupling $\Gamma=5.0$, the opposite charge distribution predicted by our formalism still exhibits fair agreement with MC data, but the theory underestimates the non-monotonic like-charge pair distribution. It is noteworthy that the corresponding departure of the SDCH prediction from the MC result is very similar to that previously observed by us for 3D electrolytes at low temperatures~\cite{Buyuk2026}.
\begin{figure}
\includegraphics[width=1\linewidth]{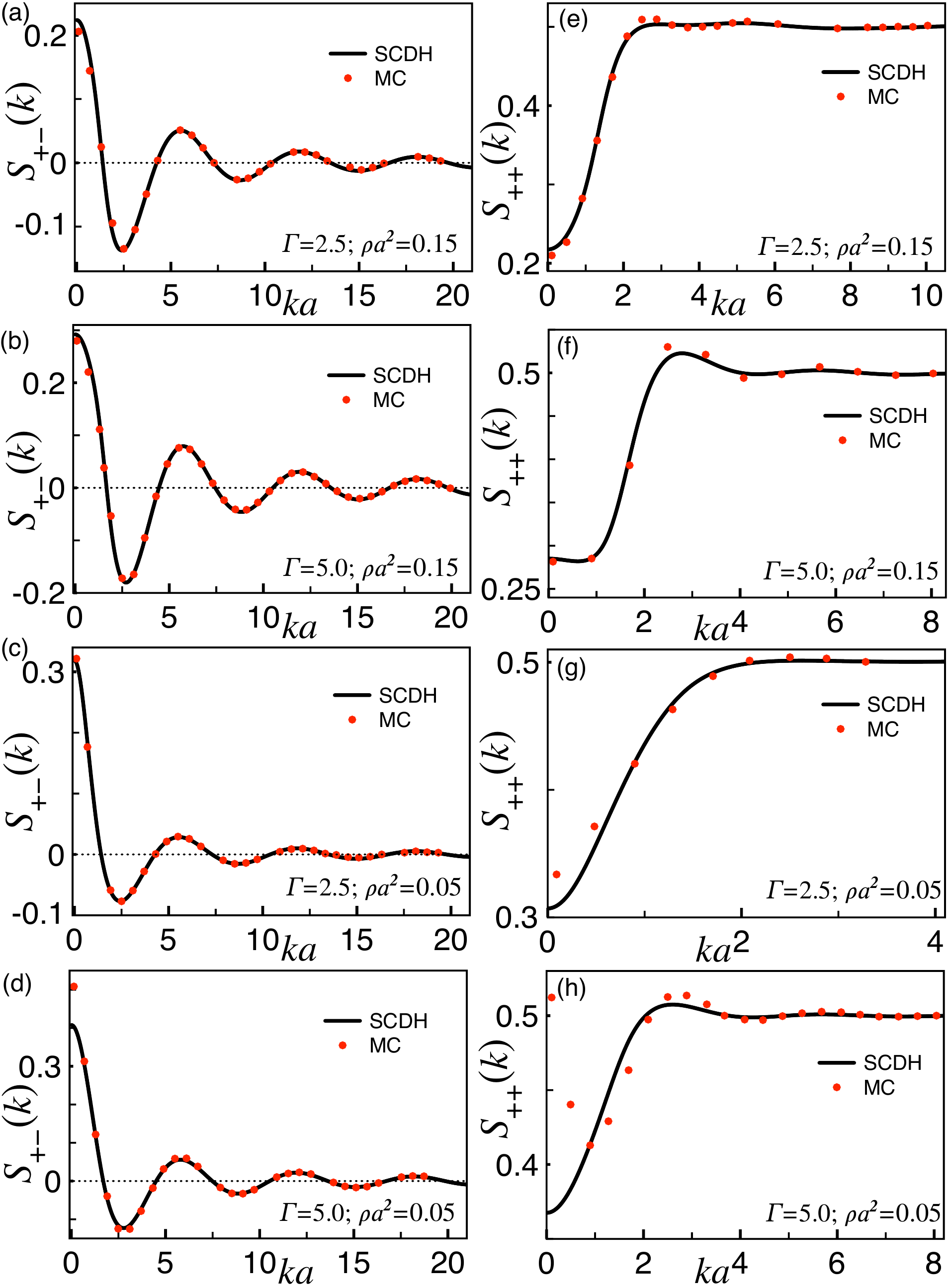}
\caption{(Color online) Structure factors. Symbols: MC data from Fig. 5 of Ref.~\cite{Lomba}. Solid curves: SCDH prediction from Eqs.~(\ref{eq64}) and~(\ref{eq69}). The values of the coupling parameter and the reduced density are indicated in the legends.}
\label{fig2}
\end{figure}

In the IR regime $ka\lesssim1$, the structure factors obtained from the FT of the pair distributions can provide accurate information on the large-distance behaviour of these distributions whose exponential decay complicates the real-space analysis of their long-range tail. Thus, in order to explain the lower accuracy of our approach in dilute solutions at substantially high electrostatic coupling, we consider now the partial structure factors defined as
\be
\label{eq67}
S_{ij}(q)=x_i\delta_{ij}+x_ix_j\rho a^2\bar{H}_{ij}(q),
\ee
where the mole fraction of the species $i$ reads $x_i=n_i/n_{\rm tot}$, and the FT of the correlation function~(\ref{eq64}) is
\be\label{eq68}
\bar{H}_{ij}(q)=-\frac{2\pi}{q}{\rm J}_1(q)+2\pi\int_1^\infty\mathrm{d}uu\;{\rm J}_0(qu)H_{ij}(u).
\ee
For 1:1 solutions with $x_i=1/2$, Eq.~(\ref{eq67}) reduces to
\be
\label{eq69}
S_{ij}(q)=\frac{1}{2}\delta_{ij}+\frac{1}{4}\rho a^2\bar{H}_{ij}(q).
\ee

Fig.~\ref{fig2} shows that  the agreement between the structure factors~(\ref{eq69}) obtained from the present formalism and the MC simulations are consistent with the radial distribution plots of Fig.~\ref{fig1}. First, one notes the SCDH theory can accurately capture the oscillatory sign inversions of the opposite-charge structure factors associated with the ion pairs (left panels) up to the intermediate coupling $\Gamma=5.0$ at the distinct density values $\rho a^2=0.05$ and $0.15$. Therein, the only significant discrepancy occurs in the IR limit ($k\to0$) of the highest coupling parameter $\Gamma=5.0$ and the dilute concentration $\rho a^2=0.05$.

Then, the inspection of Figs.~\ref{fig2}(e) and (f) indicates that at the average density $\rho a^2=0.15$ of the intermediate coupling regime $\Gamma\lesssim5.0$, the like-charge structure factors of the SCDH formalism agree fairly well with the MC simulations over the whole spectrum. Fig.~\ref{fig2}(g) shows that the deviation of the SCDH result from the MC data emerges indeed in the IR regime $ka\lesssim1$ of the reduced ion density $\rho a^2=0.05$ and moderate coupling $\Gamma=2.5$  where ionic cluster formation sets in~\cite{Lomba}.  Finally, in Figs.~\ref{fig2}(h) corresponding to the same dilute density but higher coupling $\Gamma=5.0$ characterized by enhanced ion clustering, the underestimation of the structure factor by the SCDH formalism at short wavelengths is significantly amplified. This local failure of the theory corresponding to the underestimation of the large-distance branch of the pair distribution functions $g_{ij}(r)$ implies that the SCDH approach overestimates the long-range screening of ion correlations.  These points indicate that in dilute electrolytes with substantially high electrostatic coupling, the deterioration of the quantitative accuracy of the SCDH theory  originates from its underestimation of the ionic clusters emerging close to the critical BKT line. 

\subsection{Thermodynamic functions}

In this part, we compare the thermodynamic functions of the 2D liquid obtained from the SCDH formalism with the numerical simulations of Refs.~\cite{Cailloi,Lomba,Aupic}

\subsubsection{Excess energy}

In Appendix~\ref{apen}, we report the calculation of the excess energy for a $d$- dimensional liquid. Therein, it is shown that in the case of a two-dimensional 1:1 electrolyte, the excess energy density per particle becomes
\be
\label{eq70}
\frac{\beta u_{\rm ex}}{2n_+}=\pi\Gamma n_+\int_a^\infty\mathrm{d}rr\ln\left(\frac{r}{a}\right)\left[H_{+-}(r)-H_{++}(r)\right].
\ee
In terms of the rescaled variables introduced in Sec.~\ref{resdim}, Eq.~(\ref{eq70}) takes the dimensionless form
\be
\label{eq71}
\frac{\beta u_{\rm ex}}{\rho}=\frac{\pi\Gamma}{2}\rho a^2\int_1^\infty\mathrm{d}uu\ln\left(u\right)\left[H_{+-}(u)-H_{++}(u)\right].
\ee

\begin{figure}
\includegraphics[width=1\linewidth]{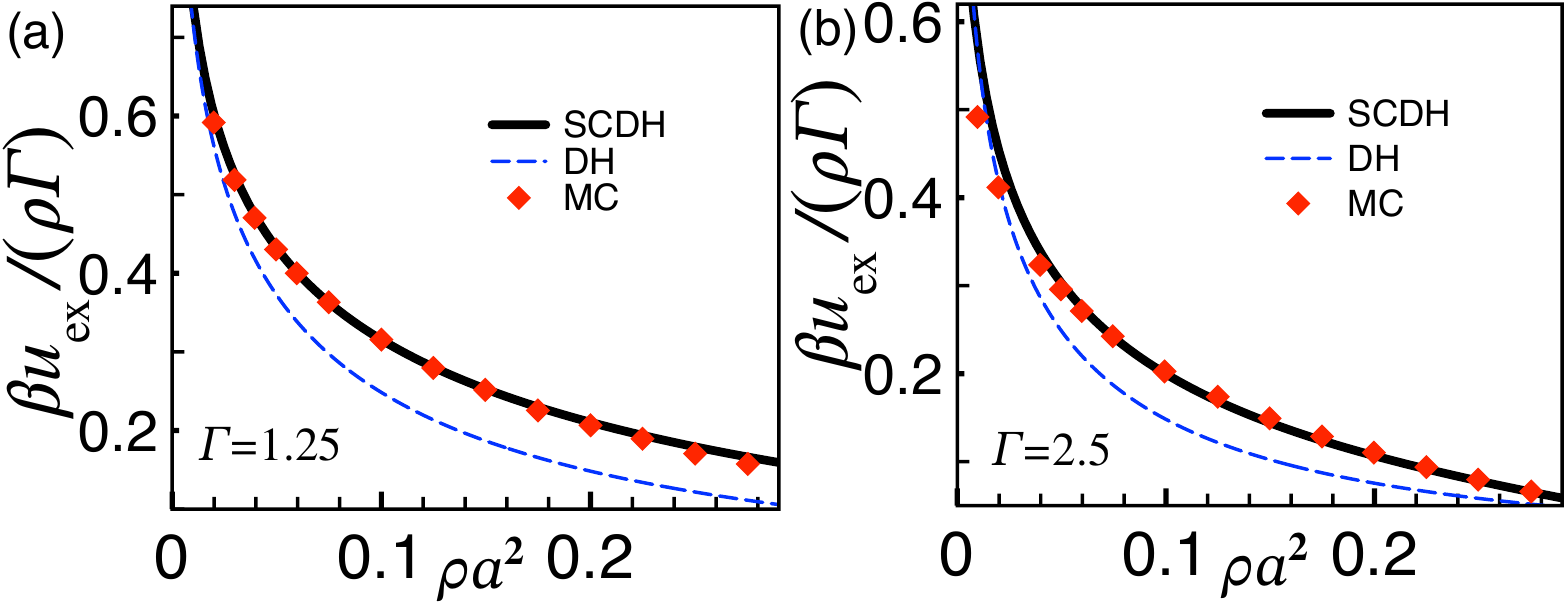}
\caption{(Color online) Excess energy density rescaled by the coupling parameter against the dimensionless density. Solid and dashed curves: SCDH and DH predictions from Eqs.~(\ref{eq71}) and~(\ref{eq73}), respectively. Symbols: MC data from (a) Fig. 6(a) and (b) Fig. 6(c) of Ref.~\cite{Aupic}. The coupling constants are indicated in the legends.}
\label{fig3}
\end{figure}
In Fig.~\ref{fig3}, we compare the internal energy~(\ref{eq71}) obtained from the SCDH theory with MC simulations. It is shown that at the coupling constant $\Gamma=1.25$, the formalism can accurately reproduce the density dependence of the excess energy up to $\rho a^2\approx0.3$. At the higher coupling $\Gamma=2.5$, the theory remains accurate in the ionic concentration regime $\rho a^2\gtrsim0.05$. At lower densities characterized by sizeable ion clustering, the formalism underestimating the ionic pair formation overestimates the charge strength of the liquid and its excess energy.

\begin{figure}
\includegraphics[width=1\linewidth]{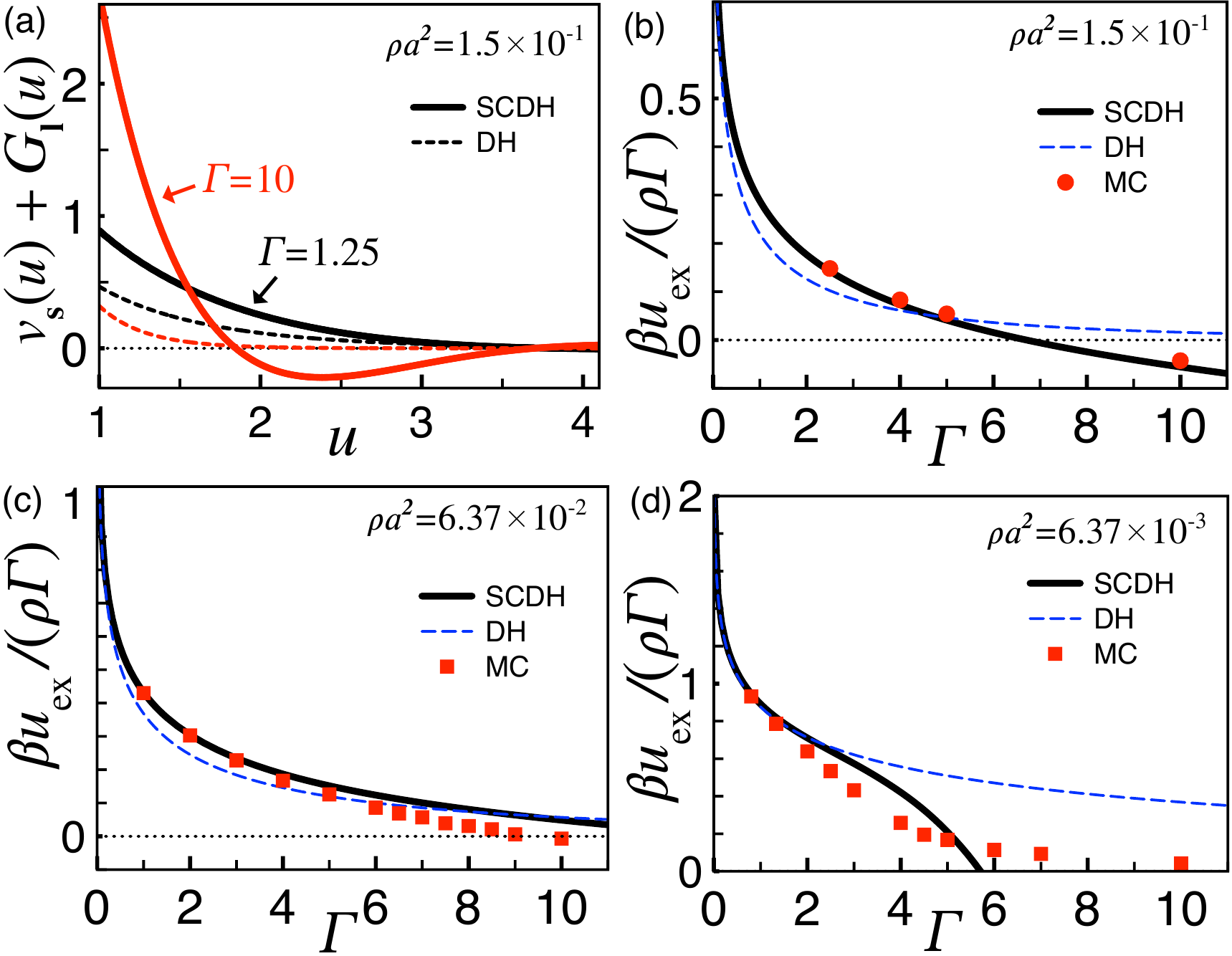}
\caption{(Color online) (a) Interionic potential of the Mayer function~(\ref{eq60}) (solid curves) and its DH limit~(\ref{eq72}) with $v\s(u)=0$ (dashed curves) at the density $\rho a^2=0.15$. The coupling parameter is $\Gamma=1.25$ (black) and $\Gamma=10$ (red). (b)-(d) Rescaled energy density against the coupling parameter $\Gamma$ at the reduced ion densities indicated in the legends.  Solid and dashed curves: SCDH and DH predictions from Eqs.~(\ref{eq71}) and~(\ref{eq73}), respectively. Symbols: MC data from (b) Fig. 8 of Ref.~\cite{Lomba} and (c)-(d) Table I of Ref.~\cite{Cailloi}.}
\label{fig4}
\end{figure}

We consider now the WC gaussian limit of the SCDH formalism and derive a simple DH-like closed-form expression for the internal energy. To this aim, we first set $t=0$ in Eq.~(\ref{eq54}) to obtain $H_{ij}(r)\simeq h_{ij}(r)$. Then, setting the splitting length to zero, i.e. $\sigma=0$, the FT of the filter function~(\ref{eq59}) simplifies to $S(q)=1$. Consequently, the short range potential~(\ref{eq57}) vanishes, i.e. $v\s(r)=0$, and the long-range potential~(\ref{eq58}) becomes 
\be\label{eq72}
G\lo(u)\approx\Gamma\;{\rm K}_0({\bar \kappa}_0 u), 
\ee
where ${\rm K}_0(x)$ stands for the modified Bessel function of the second kind~\cite{math}. Expanding now the Mayer function~(\ref{eq60}) with the potential~(\ref{eq72}) in terms of the coupling parameter $\Gamma$, and substituting the resulting approximation for the correlation function into Eq.~(\ref{eq71}), one obtains
\be\label{eq73}
\frac{\beta u_{\rm ex}}{\rho}=\frac{\Gamma}{2}{\rm K}_0({\bar \kappa}_0)+O\left(\Gamma^3,t\right).
\ee

In Fig.~\ref{fig3}, the DH-level excess energy~(\ref{eq73}) is displayed by the dashed blue line. One notes that even at the moderate coupling $\Gamma=1.25$, the validity of the DH approximation is limited to the dilute density range $\rho d^2\lesssim0.02$. At the larger coupling $\Gamma=2.5$, the DH prediction disagrees with the MC data at all concentrations.

In order to elucidate the origin of the improved accuracy provided by the SCDH formalism over the DH theory, in Fig.~\ref{fig4}(a), we compare the interionic potential of the Mayer function~(\ref{eq60}) and its DH limit~(\ref{eq72}). At this point, we note that within the framework of the SCDH formalism and the MC simulations, the presence of an HC prevents the ionic atmosphere from penetrating the hard disk surrounding each ion and thus reduces the screening experienced by the electrostatic potential of two interacting ions. In Fig.~\ref{fig4}(a), the comparison of the solid and dashed black curves indicates that at the moderate coupling $\Gamma=1.25$, the DH theory ignoring the impenetrable ionic core overestimates this screening and thus underestimates the interionic potential. In the large density regime of Fig.~\ref{fig3}, this leads to the underestimation of the electrostatic energy by the DH prediction.

Hence, the improved precision of the SCDH formalism with respect to the DH approach stems from its ability to incorporate the non-uniform screening experienced by the HC ions. Fig.~\ref{fig4}(a) shows that upon the rise of the coupling constant to $\Gamma=10$, the resulting contrast between the DH and SCDH theories becomes more pronounced. Namely, while the increase of $\Gamma$ reduces the DH potential experiencing homogeneous shielding, the SCDH potential accounting for the charge screening deficiency close to the HC volume is strongly enhanced in the corresponding region. However, outside this layer, the SCDH potential experiences a sign reversal originating from the charge screening excess imposed by the electroneutrality constraint compensating for the screening deficiency inside the HC surface. In Fig.~\ref{fig4}(b) displaying the  energy against the coupling parameter $\Gamma$, one sees that  due to the dominant contribution of this sign-reversed potential regime to the integral in Eq.~(\ref{eq71}), at the coupling $\Gamma\approx6$, the excess energy switches from positive to negative.

Figs.~\ref{fig4}(b)-(d) show that from intermediate to low ion densities, the accuracy of the DH-level excess energy~(\ref{eq73}) is limited to the WC regime $\Gamma\lesssim1$. One also notes that in consistency with the pair distributions of Fig.~\ref{fig1}, the upper coupling constant marking the validity regime of the SCDH theory decreases with the salt density. Namely, in the intermediate density regime $\rho a^2=0.15$, the SCDH result exhibits reasonable agreement with the MC data up to $\Gamma\approx10$. Then, at the lower density $\rho a^2=0.0637$, the theory remains accurate only within the reduced coupling range $\Gamma\lesssim 5$. Nevertheless, our approach can still reproduce qualitatively the trend of the MC result up to $\Gamma\sim10$.  Finally, in the strictly dilute salt regime of Fig.~\ref{fig4}(d) where one approaches the BKT line, the validity domain of the SCDH formalism shrinks to $\Gamma\lesssim 2$. Hence, at dilute densities, our formalism breaks down before reaching the coupling regime $\Gamma\ce>4$ where the conductor-insulator transition of the two-dimensional HC charges is expected to occur~\cite{Lomba}.

\subsubsection{Specific Heat}

\begin{figure}
\includegraphics[width=1\linewidth]{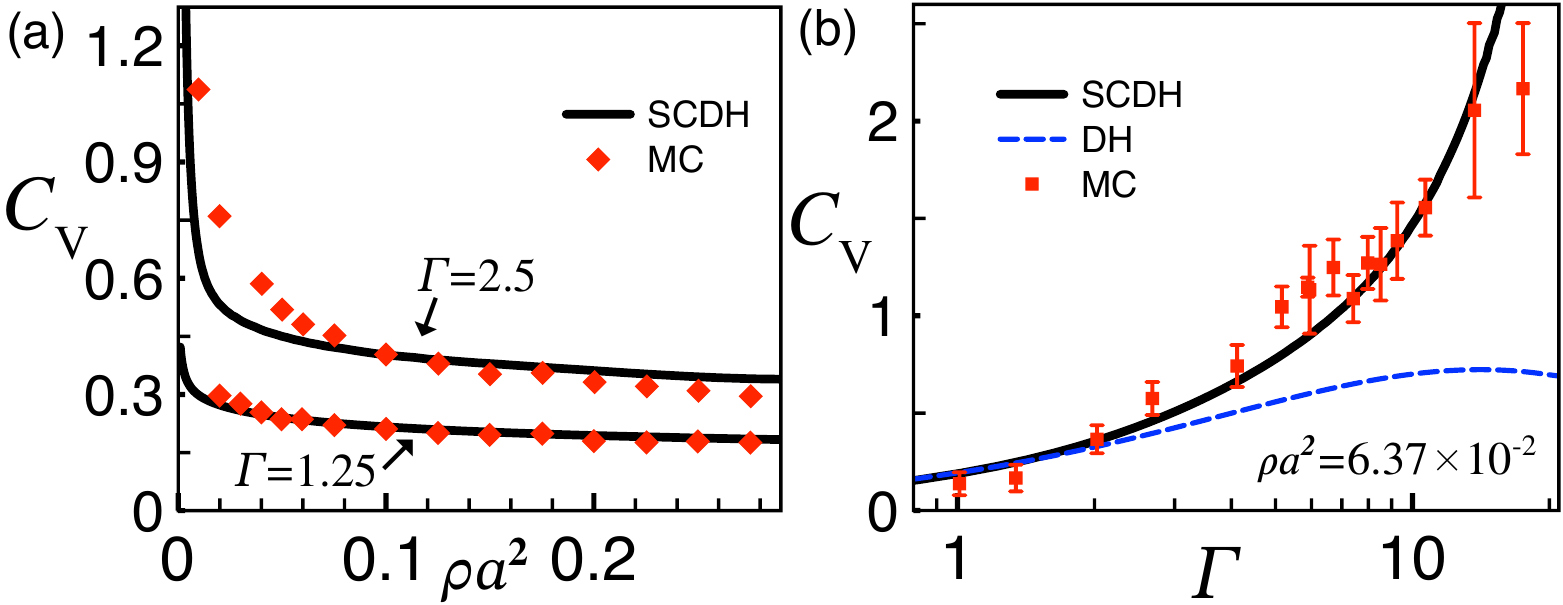}
\caption{(Color online) Specific heat against (a) dimensionless density and (b) coupling parameter.  Solid curves: SCDH prediction~(\ref{eq74}). Dashed curve in (b): DH prediction~(\ref{eq75}). Symbols: MC data (a) from Fig. 6(b) and Fig. 6(d) of Ref.~\cite{Aupic}, and (b) from Fig. 6 of Ref.~\cite{Cailloi}.}
\label{fig5}
\end{figure}
In this part, we calculate the specific heat of the 2D Coulomb liquid. The dimensionless specific heat is defined in terms of the internal energy as
\be\label{eq73II}
C_{\rm V}=\frac{1}{ k_{\rm B}}\frac{d\left(u_{\rm ex}/\rho\right)}{dT}.
\ee
Substituting the excess energy~(\ref{eq71}) into the definition~(\ref{eq73II}), and taking into account the identity $d\Gamma/dT=-\Gamma/T$, the specific heat takes the form
\be\label{eq74}
C_{\rm V}=-\frac{\pi}{2}\rho a^2\Gamma^2\int_1^\infty\mathrm{d}uu\ln\left(u\right)\frac{d}{d\Gamma}\left[H_{+-}(u)-H_{++}(u)\right].
\ee
In Eq.~(\ref{eq74}), the derivative inside the integral can be evaluated numerically  by the finite difference method. Finally, the DH limit of the specific heat~(\ref{eq74}) follows from Eqs.~(\ref{eq73}) and~(\ref{eq73II}) as
\be\label{eq75}
C_{\rm V}=\frac{\Gamma}{4}{\bar \kappa}_0{\rm K}_1({\bar \kappa}_0)+O\left(\Gamma^3,t\right).
\ee

In Fig.~\ref{fig5}(a), we plotted the specific heat~(\ref{eq74}) versus the charge density at the coupling parameters of Fig.~\ref{fig3}. One notes that the agreement of the specific heat curves with the MC data is mostly consistent with the energy plots in Fig.~\ref{fig3}. More precisely, at the coupling constant $\Gamma=1.25$, the SCDH formalism can accurately reproduce the specific heat over the entire density range $\rho a^2\lesssim0.3$. At the larger coupling $\Gamma=2.5$, the accuracy of the SCDH approach is acceptable at large densities  $\rho a^2\gtrsim0.05$, but the rapid drop of the specific heat at lower densities can be reproduced by our formalism only qualitatively. 

Finally, in Fig.~\ref{fig5}(b), we display the dependence of the specific heat on the coupling parameter at the ion density of Fig.~\ref{fig4}(b). One notes that within the fluctuations of the MC data, the general trend of the specific heat can be reproduced by the SCDH formalism up to $\Gamma\approx10$. The plot also shows that the validity of the DH result~(\ref{eq75}) is limited to $\Gamma\lesssim2$. Beyond that coupling strength, the DH approximation overestimating the charge screening by the ionic atmosphere underestimates the specific heat.

\section{Conclusions}

A self-consistent field theory of bulk electrolytes incorporating ionic HC size and electrostatic interactions on an equal footing has been applied to the 2D Coulomb liquid. Via the systematic comparison of the radial distributions, the structure factors, and the thermodynamic functions obtained from the SCDH approach with MC simulation data from the literature, we identified the validity domain of the formalism in terms of the electrostatic coupling strength and the ion density. This comparative study shows that the main accomplishment of the 2D SCDH formalism is the accurate characterization of the thermodynamics of HC ions from weak to intermediate coupling regime at average salt densities. The analysis of the many-body-dressed interionic potentials indicates that the improved accuracy of the SCDH theory with respect to the WC-level DH approach is mainly due the ability of the former to account for the non-uniform screening  of electrostatic interactions caused by the impenetrability of the hard disks by their ionic atmosphere.

In the average density regime $0.15\gtrsim\rho a^2>0.05$, the pair distributions and the structure factors of the numerical simulations are accurately reproduced by the SCDH approach up to intermediate coupling strengths $\Gamma\approx5$. We found that the upper coupling constant marking the validity limit of the formalism drops with the ion density. Indeed, the analysis of the structure factors showed that as one moves into the dilute salt regime $\rho a^2\lesssim0.05$ where sizeable ion clustering occurs~\cite{Lomba}, our computational approach overestimates the long-range screening of the pair distribution functions and thus the charge strength of the electrolyte. This indicates that the shrinking of the validity domain of the SCDH formalism upon salt decrement stems from the underestimation of the ionic pair formation continuously intensified towards the critical BKT line. 

The comparison of the excess energy and specific heat curves with MC results provided a more precise identification of the validity regime of our approach. Namely, we showed that within  the entire salt density range $\rho a^2\lesssim0.3$, the present formalism can accurately predict the thermodynamic functions up to moderate couplings $\Gamma\approx2$. However, at intermediate couplings $\Gamma\gtrsim2$, the quantitative accuracy of the theory maintained at average densities $\rho a^2\gtrsim0.05$ deteriorates in the dilute charge regime $\rho a^2<0.05$. That is, the SCDH approach breaks down below the critical coupling domain $\Gamma\ce>4$ where the BKT transition of dilute hard disks occurs~\cite{Lomba}. This indicates that accessing the 2D conductor-insulator transition via the present formalism will require the extension of the SC scheme at the basis our approach at least up to the next cumulant order.

It is noteworthy that owing to the incorporation of the electrostatic and HC interactions into the partition function on an equal level, the SCDH approach has potential for numerous improvements. Indeed, the corresponding theoretical framework is adequate for various systematic upgrades such as the extension of the SC cumulant approximation underlying our formalism to higher orders, the consideration of generalized variational splitting schemes, the inclusion of more realistic charge structures and ion specificity, and the generalization of the theory to nanoconfined fluids. Extensions of the present formalism along these lines will be considered in upcoming works.

\smallskip
\appendix

\section{Excess Energy in $d$-dimensions}
\label{apen}

\begin{widetext}
We derive here the excess energy in a $d$-dimensional space of volume $V_d$ and solid angle $\Omega_d$~\cite{Hansen,Buyuk2024,Soft2024}. The excess energy is defined as the GC average of the total energy in the Boltzmann distribution of Eq.~(\ref{eq4}) without the self-energy component, i.e. $\beta U_{\rm ex}=\lan\beta(E\ce+E\h-E\s)\ran\G$, or
\be
\label{apeq1}
\beta U_{\rm ex}=\frac{1}{2}\int\mathrm{d}^d\br\mathrm{d}^d\br'\left[v\ce(\br,\br')\lan\hc(\br)\hc(\br')\ran\G+v\h(\br,\br')\lan\hh(\br)\hh(\br')\ran\G\right]-\beta E\s.
\ee
Inserting into Eq.~(\ref{apeq1}) the density operators in Eq.~(\ref{eq8}) and the self-energy~(\ref{eq7}), one obtains
\be
\label{apeq2}
\beta U_{\rm ex}=\frac{1}{2}\sum_{i=1}^p\sum_{j=1}^p\int\mathrm{d}^d\br\mathrm{d}^d\br'\left[q_iq_j v\ce(\br,\br')+v\h(\br,\br')\right]\left[\lan \hn_i(\br)\hn_j(\br')\ran\G-n_i\delta_{ij}\delta^d(\br-\br')\right]
\ee
At this point, using the definition of the pair distribution~(\ref{eq28}), Eq.~(\ref{apeq2}) can be expressed as
\bea
\label{apeq3}
\beta U_{\rm ex}&=&\frac{1}{2}\sum_{i=1}^p\sum_{j=1}^pn_in_j\int\mathrm{d}^d\br\mathrm{d}^d\br'\left[q_iq_j v\ce(\br,\br')+v\h(\br,\br')\right]g_{ij}(\br,\br')\\
&=&\frac{\Omega_dV_d}{2}\sum_{i=1}^p\sum_{j=1}^pn_in_j\int_0^\infty\mathrm{d}rr^{d-1}\left[q_iq_j v\ce(r)+v\h(r)\right]g_{ij}(r).
\eea
\end{widetext}
In order to derive the second equality of Eq.~(\ref{apeq3}), we exploited the translational and spherical symmetries implying $v\ce\left(\br,\br'\right)=v\ce\left(||\br-\br'||\right)$, $v\h\left(\br,\br'\right)=v\h\left(||\br-\br'||\right)$, and $g_{ij}\left(\br,\br'\right)=g_{ij}\left(||\br-\br'||\right)$. Taking now into account the HC cut-off of the pair distribution function, i.e. $g_{ij}(r)\propto\theta(r-a)$, one finds that the contribution from the HC potential to the integral in Eq.~(\ref{apeq3}) vanishes, i.e.
\be
\label{apeq4}
\int_0^\infty\mathrm{d}rr^{d-1}v\h(r)g_{ij}(r)=\int_a^\infty\mathrm{d}rr^{d-1}v\h(r)g_{ij}(r)=0,
\ee
where we accounted for the cancellation of the HC potential at $r>a$. Hence, taking into account as well the global electroneutrality condition~(\ref{eq32}) and the definition of the total correlation function~(\ref{eq34II}), the excess energy density $u_{\rm ex}=U_{\rm ex}/V_d$ simplifies to
\be
\label{apeq4II}
\beta u_{\rm ex}=\frac{\Omega_d}{2}\sum_{i=1}^p\sum_{j=1}^pn_in_jq_iq_j \int_a^\infty\mathrm{d}rr^{d-1}v\ce(r)H_{ij}(r).
\ee
Finally, setting $d=2$ and accounting for the definition of the 2D Coulomb potential~(\ref{eq2}), the 2D excess energy density follows from Eq.~(\ref{apeq4II}) as
\be
\label{apeq5}
\beta u_{\rm ex}=-\pi\Gamma\sum_{i=1}^p\sum_{j=1}^pn_in_jq_iq_j\int_a^\infty\mathrm{d}rr\ln\left(\frac{r}{a}\right)H_{ij}(r).
\ee
In the case of the 1:1 electrolyte investigated in the present work, considering that $p=2$, $n_+=n_-$, and $q_\pm=\pm1$, the excess energy density per particle~(\ref{apeq5}) reduces to
\be
\label{apeq6}
\frac{\beta u_{\rm ex}}{2n_+}=\pi\Gamma n_+\int_a^\infty\mathrm{d}rr\ln\left(\frac{r}{a}\right)\left[H_{+-}(r)-H_{++}(r)\right].
\ee


\begin{thebibliography}{99}
\bibitem {rem1} In Eq.~(\ref{eq19}), the addition of derivatives higher than eight order does not induce any significant improvement of the agreement with MC simulations.
\bibitem {rem2} The simplification that consists in limiting ourselves to the gaussian-level incorporation of long-range ion correlations is motivated by the fact that within the validity regime of our formalism closely identified in this article, the beyond gaussian-level kernel correction of order $O(t)$ brings an insignificant contribution to thermodynamic functions.
\end{thebibliography}
\end{document}